\documentclass[11pt]{article}
\usepackage{natbib}
\usepackage{float}
\usepackage{url}
\usepackage{amsmath}
\usepackage{booktabs}
\usepackage{graphicx}
\usepackage{float}
\usepackage[margin=1in]{geometry}
\usepackage{smartdiagram}
\usesmartdiagramlibrary{additions}
\title{Causal Inference under Interference with Learned Exposure Mappings}

\author{
Cong Cao, PhD\\
Department of Biostatistics\\
Yale School of Public Health\\
Yale University\\
New Haven, CT 06520, USA\\
\texttt{cong.cao@yale.edu}
}

\date{}
\begin{document}

\maketitle

\begin{abstract}
Exposure mappings are often assumed to be known in causal spillover analyses. In environmental settings, however, they are typically induced by transport processes that are not directly observed and must instead be learned from pollution data. We study how uncertainty in learned transport processes propagates into exposure mappings and downstream spillover inference under interference. We compare mechanistic transport models with modern operator-learning approaches, including PDE, PINO, FNO, and GeoPT, using both simulation studies and an empirical analysis of California PM$_{2.5}$ data. In simulations, all four transport models achieved nearly identical pollution prediction accuracy, yet estimated spillover effects ranged from 1.78 to 2.27. Models that more accurately recovered the induced exposure mapping also produced spillover estimates closer to the true effect. Disagreement was modest for regional interventions but substantially larger for localized point-source interventions. The California analysis showed the same pattern: competing transport models produced similar predictions of observed PM${2.5}$ concentrations while implying different spillover effects under hypothetical pollution-control interventions. Our findings suggest that predictive agreement alone is insufficient for reliable causal inference when exposure mappings are learned rather than directly observed.

\end{abstract}


\section*{Introduction}

Causal inference under interference studies settings where treatment assigned to one unit can affect outcomes in other units. A central component of this framework is the exposure mapping, which summarizes how treatments received by other units determine the effective exposure of each unit \cite{hudgens2008toward,aronow2012estimating,ogburn2014causal,tchetgen2012causal,savje2021average}. Most existing methods assume that this exposure mapping is known in advance and can be specified using information such as geographic distance, network connections, or predefined neighborhood structures.

In many environmental applications, however, the exposure mapping is not directly observed. Consider a policy that reduces emissions at selected locations. The resulting health effects depend not only on where emissions are reduced but also on how pollutants move through the atmosphere. Because atmospheric transport cannot be observed directly, the exposure mapping must be learned or specified from pollution data rather than treated as known.

 Recent work has proposed constructing exposure mappings using atmospheric transport models rather than fixed spatial weights \cite{wikle2024causal}. Under this framework, observed pollution concentrations are used to estimate a transport process, which in turn defines the exposure mapping used for causal analysis. Consequently, causal conclusions depend directly on assumptions about the estimated transport process.

 This raises a central question. A wide range of transport models is now available, including mechanistic PDE models, physics-informed neural operators (PINO), Fourier neural operators (FNO), and more recent foundation models such as GeoPT\cite{tran2021factorized,kovachki2023neural,wu2026geopt}. Although these approaches often achieve similar predictive accuracy, they may imply different transport operators and, consequently, different exposure mappings. More generally, predictive performance and scientific inference need not coincide \cite{breiman2001statistical,shmueli2010explain}. Models that fit the observed data equally well may nevertheless support different scientific conclusions, particularly when the quantities of interest depend on latent structures rather than directly observed outcomes. Similar predictive performance therefore does not necessarily imply similar causal conclusions about spillover effects. Although we focus on atmospheric transport, the same challenge arises whenever exposure mappings are learned from latent physical or mechanistic processes rather than directly observed.

\begin{figure}[H]
\centering
\includegraphics[width=0.90\textwidth]{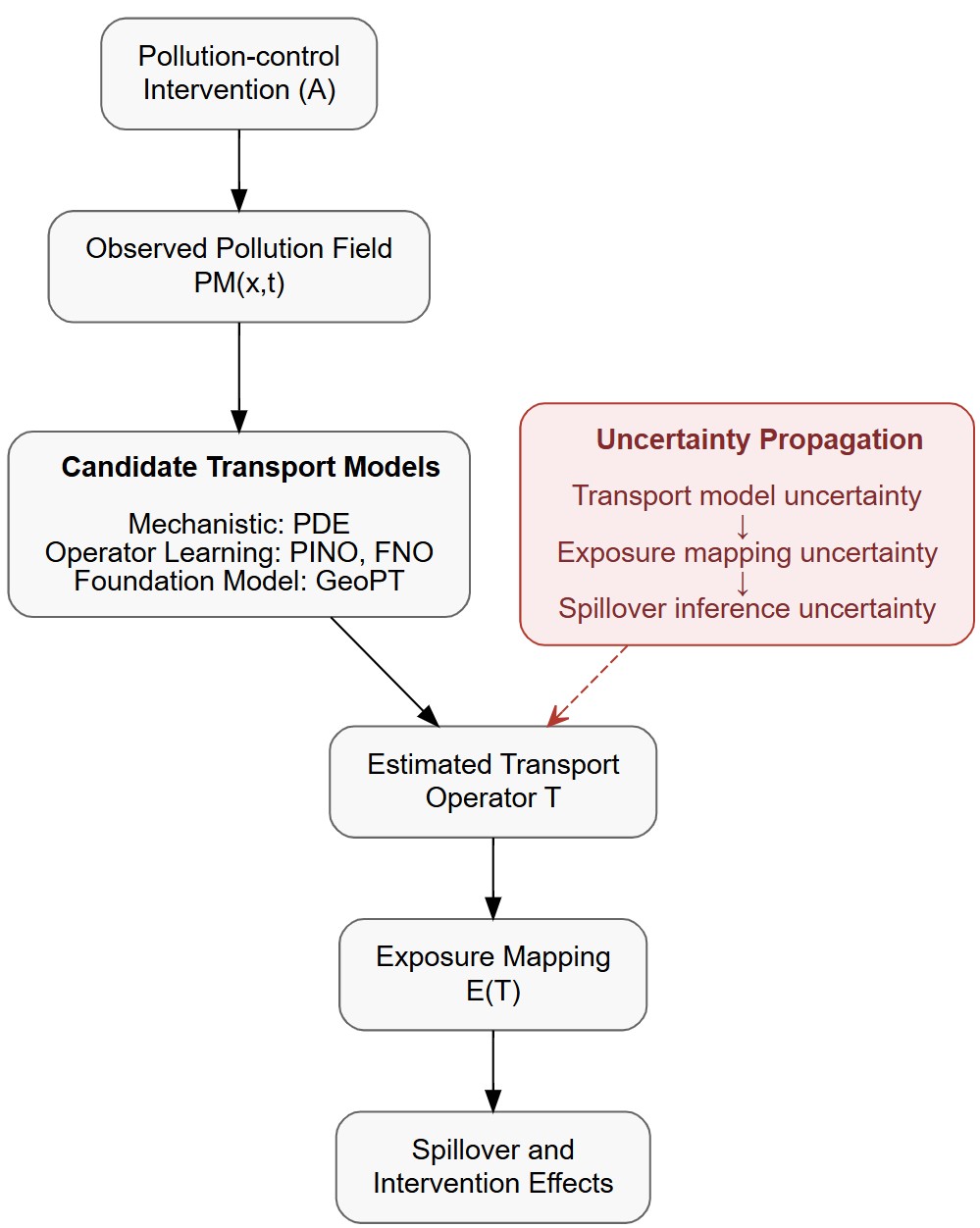}
\caption{
Uncertainty propagation from transport modeling to spillover inference.
Observed pollution data support multiple plausible transport models.
Although competing models may achieve similar predictive performance,
they can induce different exposure mappings through different estimated transport operators,
leading to different spillover estimates and potentially different causal conclusions.
}
\label{fig:framework}
\end{figure}

In this paper, we study how uncertainty in learned transport processes affects spillover inference under interference. We compare mechanistic transport models with modern operator-learning approaches, including PDE, PINO, FNO, and GeoPT, using both simulation studies and an empirical analysis of California PM$_{2.5}$ data. Rather than asking which transport model is correct, we ask whether transport models with similar predictive performance can nevertheless induce different exposure mappings and therefore different spillover conclusions. Across both simulated and empirical settings, we show that they can. Our results demonstrate that predictive agreement alone is insufficient for reliable causal inference when exposure mappings are learned rather than directly observed. Figure \ref{fig:framework} outlines the proposed framework. The subsequent simulations illustrate how uncertainty propagates into spillover inference, and the final summary figure synthesizes the main findings.
\section*{Methods}
 
\subsection*{Conceptual Setup}

We consider a pollution-control intervention $A$, observed pollution concentrations $PM$, a transport process $T$, and a health outcome $H$. Given a candidate transport process $T$, we define an exposure mapping $E(T)$ that characterizes interference across locations.

Most approaches to causal inference under interference assume that the exposure mapping is known. In environmental applications, however, the exposure mapping is typically induced by an unobserved transport process. Consider an intervention vector

\[
A=(A_1,\ldots,A_N),
\]

where \(A_i\) represents an emission reduction or pollution-control action at location \(i\). The scientific quantity of interest is not pollution itself but rather the downstream health consequences of such interventions. Because pollutants move through space, interventions applied at one location may influence health outcomes at other locations through atmospheric transport.

Let

\[
T=(T_{ij})
\]

denote a transport operator, where \(T_{ij}\) describes the contribution of emissions originating at location \(j\) to exposure at location \(i\). Under transport model \(T\), the spillover exposure induced by intervention pattern \(A\) is

\[
E(T)=TA,
\]

with unit-level exposure

\[
E_i(T)=(TA)_i.
\]

Throughout the paper, uncertainty arises because the transport operator $T$ is not observed and must be learned from pollution observations $PM$. Following the interference framework, let

\[
H_i(a,e)
\]

denote the potential health outcome at location \(i\) under direct intervention level \(a\) and spillover exposure level \(e\). Observed outcomes satisfy

\[
H_i
=
H_i\!\left(A_i,E_i(T)\right).
\]

\[
H_i
=
\beta_0
+
\beta_1A_i
+
\beta_2E_i(T)
+
\varepsilon_i,
\]

where \(\beta_2\) summarizes the relationship between spillover exposure and health outcomes under a given transport model. Although spillover coefficients are reported throughout the analysis, our primary focus is on the stability of intervention effects across observationally equivalent transport models. The specific form of the outcome model is not the focus of this paper. Rather, the goal is to examine how uncertainty in the transport process propagates through the induced exposure mapping and affects downstream spillover inference.

 In the simulation study, a latent transport operator $T^\star$ generates the pollution dynamics and the corresponding interference process. Pollution observations are used to estimate several candidate transport models. Although these models may fit the observed pollution field similarly well, they can induce different transport operators and exposure mappings. We compare the resulting spillover effect estimates across candidate models.
 
\subsection*{Simulation Study}

For each simulation replicate, a latent transport operator \(T^\star\) was used to generate pollution dynamics across space. The resulting pollution field was observed, whereas the true transport operator remained unobserved. Competing transport models, including an advection--diffusion PDE model, PINO, FNO, and GeoPT, were fitted to the same pollution observations. Each fitted model produced an estimated transport operator \(\widehat T\), which induced an exposure mapping

\[
E(\widehat T)=\widehat T A.
\]

The resulting exposure mapping was used to estimate spillover and intervention effects. We compared competing transport models in terms of pollution prediction accuracy, recovery of the exposure mapping, and the resulting causal effect estimates. We further examined how uncertainty in the learned transport process translated into variation in downstream spillover and intervention effect estimates.

\section*{Results}

We first summarize the overall performance of the competing transport models before examining where differences in spillover estimation arise.

Among the four transport models, FNO produced the smallest exposure mapping error and the spillover estimate closest to the true value. In this simulation setting, accurate recovery of the exposure mapping was more closely associated with spillover estimation than pollution prediction accuracy. Although GeoPT achieved prediction accuracy comparable to the other transport models, its exposure mapping differed more substantially from the latent transport process, leading to larger deviations in spillover estimation.
\begin{table}[H]
\centering
\caption{Prediction accuracy, exposure recovery, and spillover estimation under competing transport models.
}
\label{tab:main_results}
\begin{tabular}{lcccc}
\toprule
Model & PM $R^2$ & Exposure RMSE & Spillover Effect & Rank Corr \\
\midrule
PDE   & 0.975 & 0.0087 & 2.272 & 1.000 \\
PINO  & 0.972 & 0.0112 & 1.784 & 1.000 \\
FNO   & 0.975 & 0.0013 & 2.002 & 1.000 \\
GeoPT & 0.970 & 0.0082 & 1.877 & 0.995 \\
\bottomrule
\end{tabular}
\end{table}

\subsection*{Observationally Equivalent Transport Models}
\begin{figure}[H]
\centering
\includegraphics[width=0.75\textwidth]{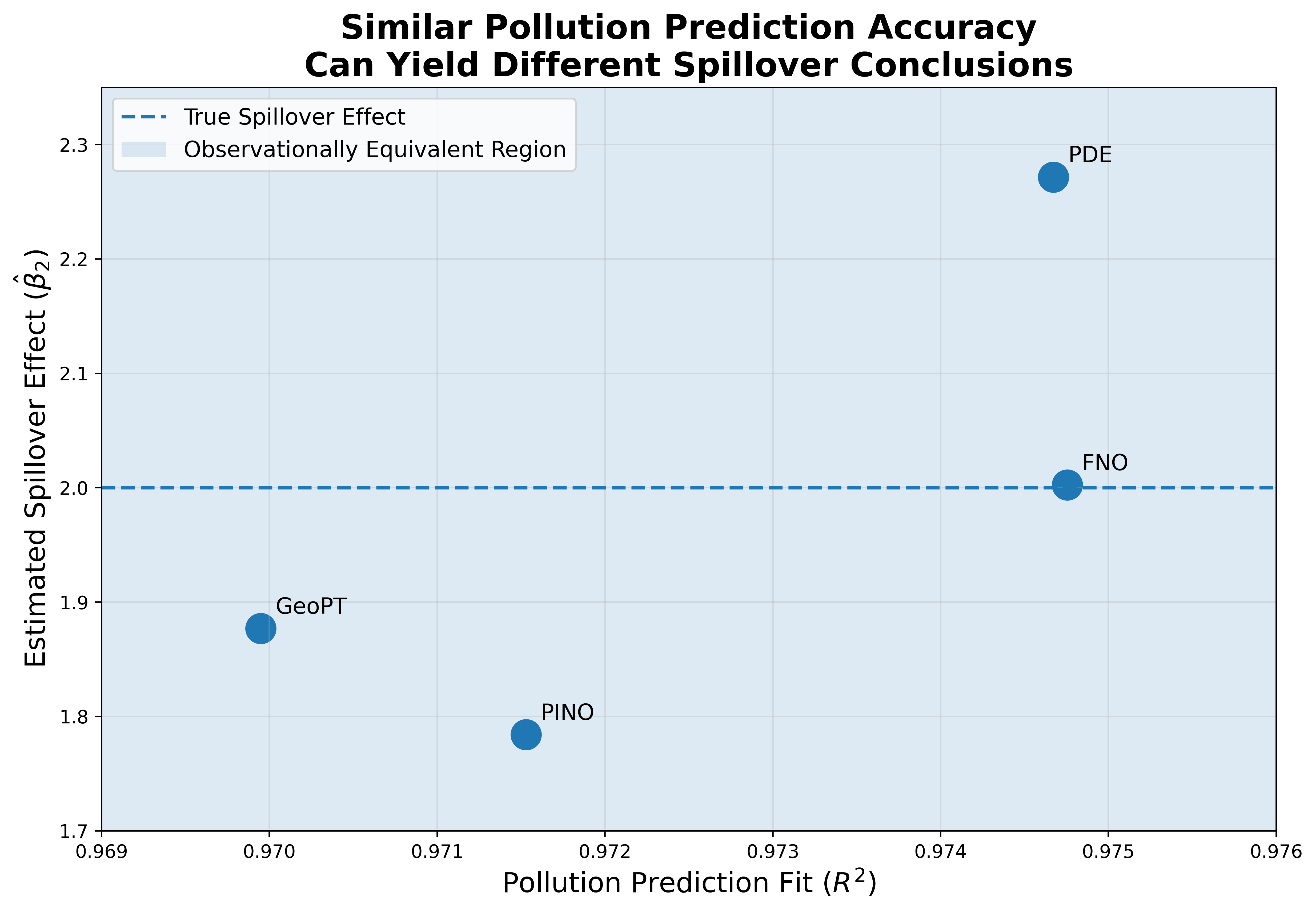}
\caption{
Observationally equivalent transport models.
Several transport models achieved similar pollution prediction accuracy while producing different spillover estimates.
}
\label{fig:obs_equivalence}
\end{figure}

We first examined whether competing transport models could reproduce the observed pollution field equally well. Despite substantial differences in model structure, all four models achieved nearly identical predictive performance. The prediction of PM \(R^2\) ranged from 0.970 to 0.975 in PDE, PINO, FNO, and GeoPT (Table~\ref{tab:main_results}). Although the models were nearly indistinguishable in terms of pollution prediction, they did not produce identical spillover estimates. Estimated spillover effects ranged from 1.78 to 2.27, with FNO recovering a value closest to the true effect. Figure~\ref{fig:obs_equivalence} illustrates that models clustered tightly with respect to PM prediction accuracy while showing substantially greater variation in spillover estimation.

\subsection*{Exposure Mapping Uncertainty}
\begin{figure}[H]
\centering
\includegraphics[width=0.75\textwidth]{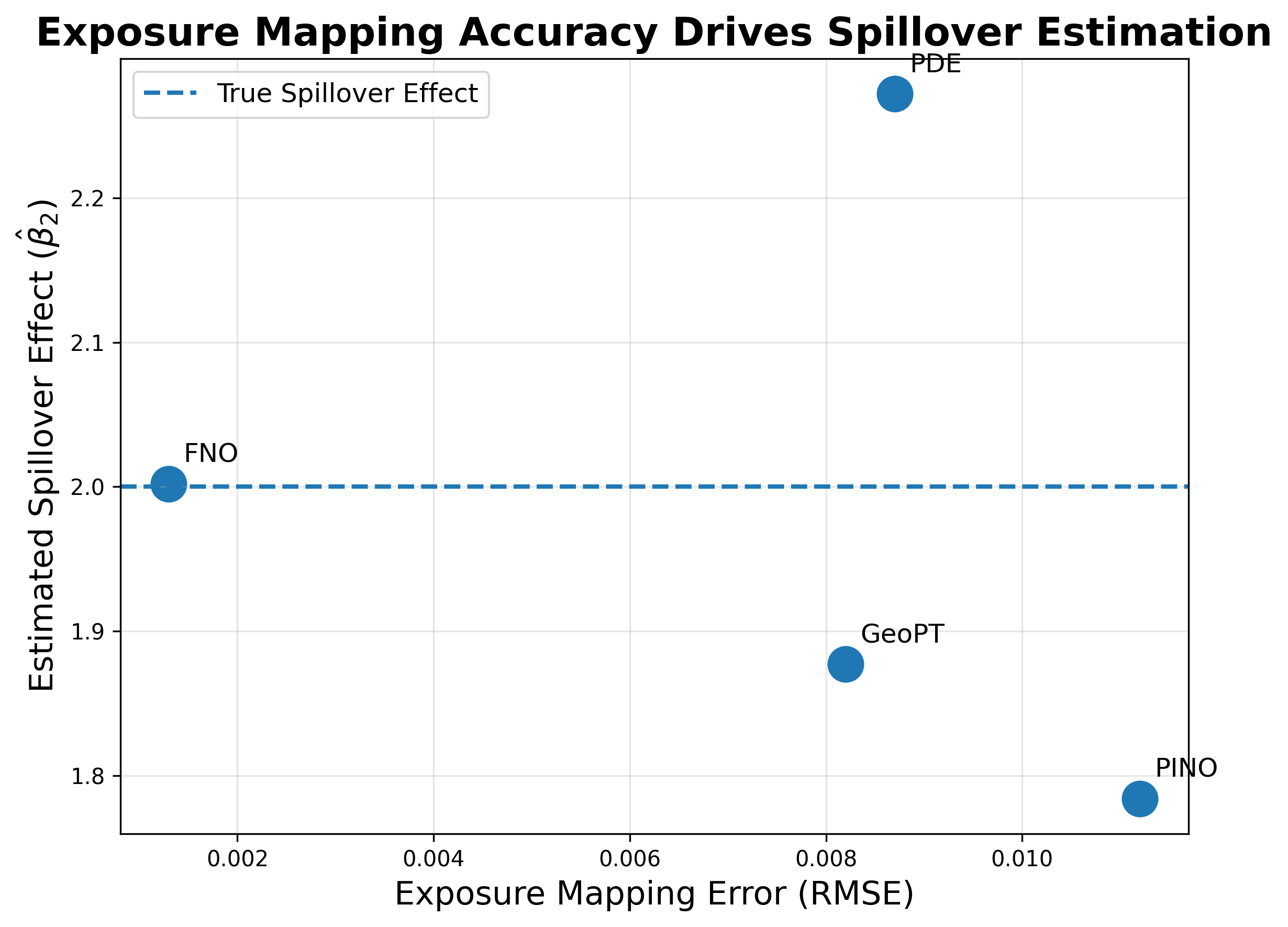}
\caption{
Exposure mapping accuracy was more strongly associated with spillover estimation than with pollution prediction accuracy. Models with larger exposure mapping errors tended to produce larger deviations in estimated spillover effects.
}
\label{fig:exposure_vs_beta}
\end{figure}

To understand where these differences came from, we next examined the exposure mappings implied by each transport model. Although PM prediction accuracy was nearly identical across models, exposure recovery accuracy varied more substantially. Exposure RMSE ranged from 0.0013 for FNO to 0.0112 for PINO (Table~\ref{tab:main_results}). Figure~\ref{fig:exposure_vs_beta} shows that models with more accurate exposure mappings generally produced spillover estimates closer to the true value. The true spillover coefficient was 2.0. FNO recovered a value of 2.00 almost exactly, whereas PINO and GeoPT underestimated the spillover effect, and the PDE model slightly overestimated it. In contrast, models with larger exposure errors tended to produce larger deviations in the estimated spillover effect. Figures~\ref{fig:obs_equivalence} and \ref{fig:exposure_vs_beta} suggest that pollution prediction accuracy alone is not sufficient for reliable spillover inference. Two transport models may fit the observed pollution field equally well while implying different exposure mappings. These results suggest that recovering the exposure mapping is more important for spillover estimation than achieving marginal improvements in pollution prediction accuracy. We next investigate whether these differences translate into meaningful differences under alternative intervention strategies.

\subsection*{Consequences for Intervention Scenarios}
We next examined whether exposure mapping uncertainty affected conclusions about environmental interventions. Three intervention scenarios were considered: a regional emission reduction, a localized point-source intervention, and an upwind intervention. The consequences of exposure mapping uncertainty depended strongly on the intervention being considered. For regional interventions, estimated spillover effects ranged from 1.79 to 2.42 across transport models. For upwind interventions, the range was smaller, from 1.93 to 2.13. In contrast, point-source interventions produced substantially larger disagreement. Estimated spillover effects ranged from 0.51 under the mechanistic PDE model to 1.56 under GeoPT, corresponding to a three-fold difference in the estimated policy effect.

\begin{table}[H]
\centering
\caption{Sensitivity of spillover estimates across policy interventions.}
\label{tab:policy_range}
\begin{tabular}{lccc}
\toprule
Policy & Min $\hat{\beta}_2$ & Max $\hat{\beta}_2$ & Range \\
\midrule
Regional & 1.79 & 2.42 & 0.63 \\
Point Source & 0.51 & 1.56 & 1.05 \\
Upwind & 1.93 & 2.13 & 0.21 \\
\bottomrule
\end{tabular}
\end{table}

\begin{figure}[H]
\centering
\includegraphics[width=0.80\textwidth]{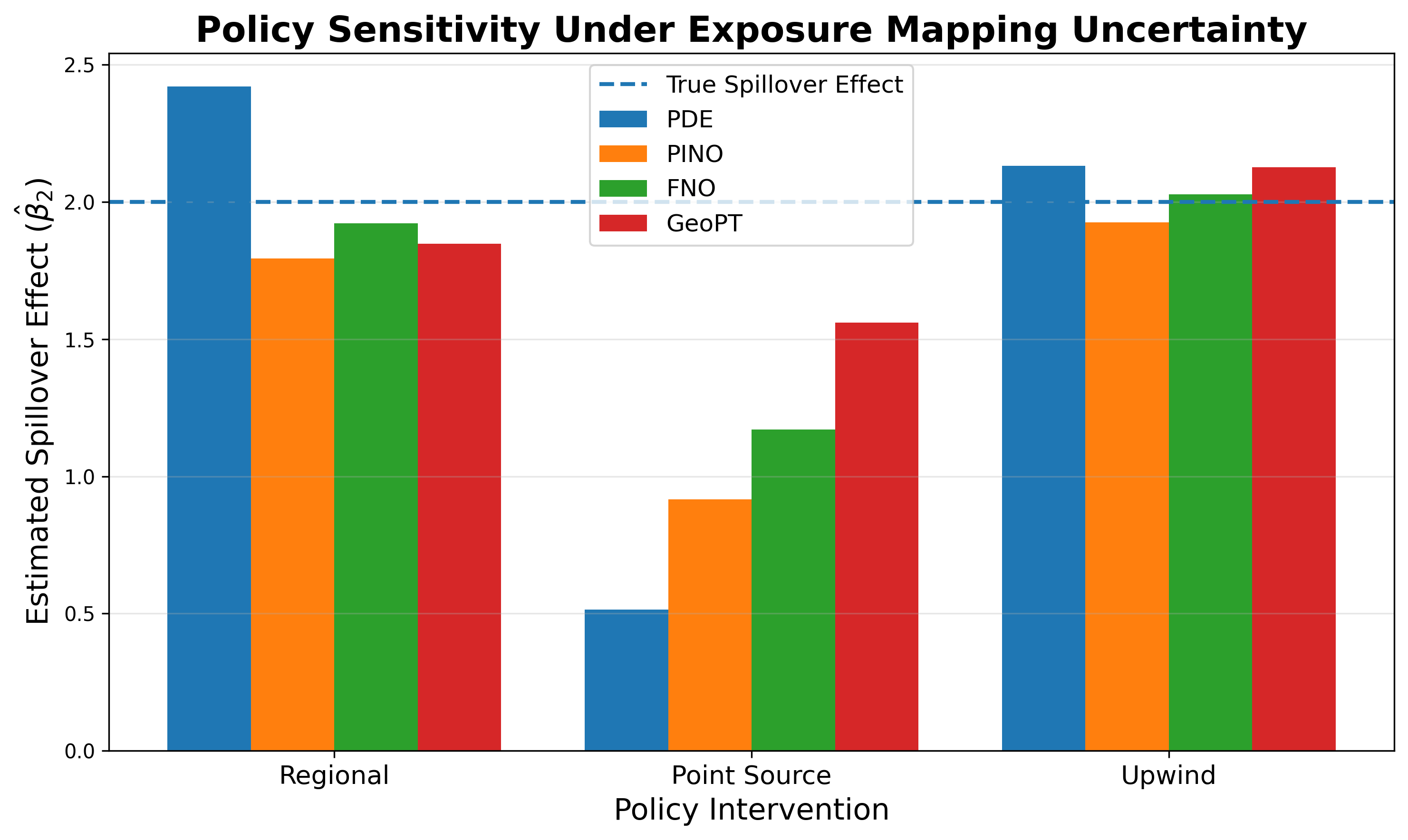}
\caption{
Estimated spillover effects under competing transport models across intervention scenarios. Differences across transport models were modest for regional and upwind interventions but substantially larger for localized point-source interventions.
}
\label{fig:policy_Surface}
\end{figure}

The contrast between Figures~\ref{fig:obs_equivalence}
and \ref{fig:exposure_vs_beta} highlights an important distinction. Models that achieved nearly identical PM prediction accuracy could nevertheless differ substantially in exposure recovery, leading to different spillover estimates. Figure~\ref{fig:policy_Surface} summarizes these results. While regional and upwind interventions produced broadly similar conclusions across transport models, localized interventions were much more sensitive to exposure mapping assumptions. Small differences in transport behavior translated into large differences in predicted spillover effects when the intervention was concentrated at a single location. The greater stability under regional interventions suggests that aggregating treatment across many locations reduces the impact of local discrepancies among competing transport models. This finding suggests that uncertainty in learned transport processes may matter most for targeted interventions, where local transport behavior has a stronger influence on downstream exposure.

Localized interventions concentrated treatment at a small number of locations. Consequently, small differences in the estimated transport process produced larger differences in spillover exposure than under regional interventions, where transport effects were averaged across many locations.

 The disagreement was not limited to the overall magnitude of the spillover effect. Although the models produced similar pollution predictions, they did not always identify the same locations as being most affected by the intervention. Exposure mapping uncertainty therefore influenced both the size and geographic distribution of estimated spillover effects.  These findings complement rather than contradict
the previous results. Exposure mapping uncertainty may substantially affect scientific understanding of spillover processes even when intervention effects remain stable.

\begin{figure}[H]
\centering
\includegraphics[width=\textwidth]{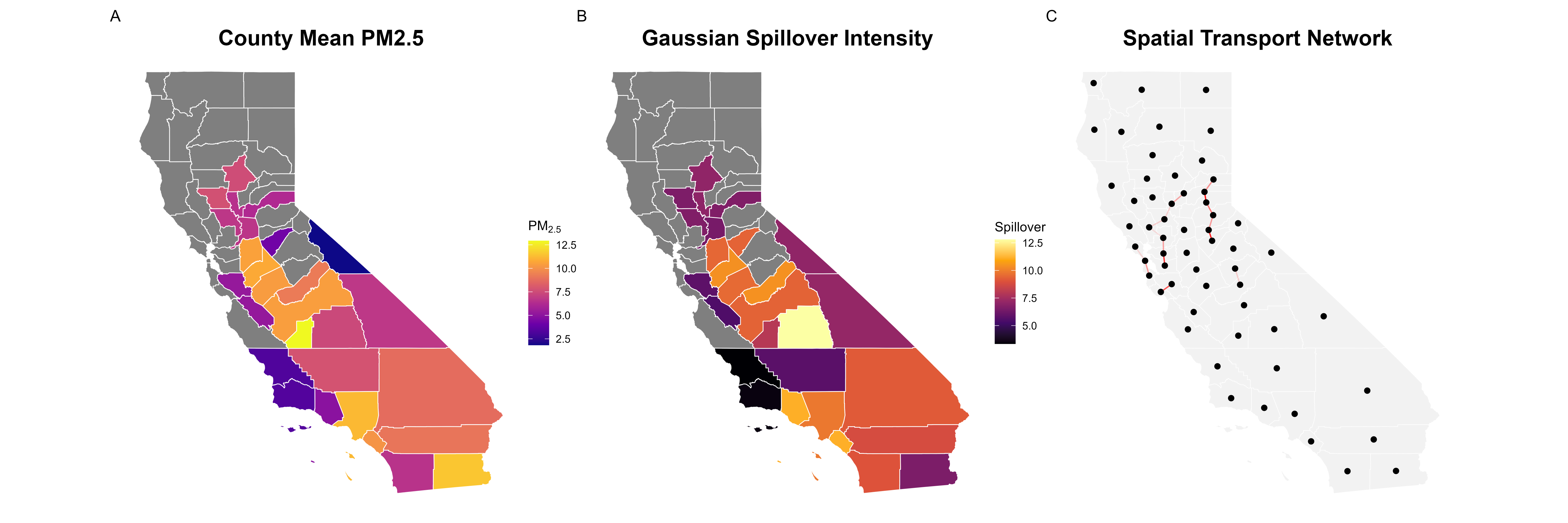}
\caption{
California illustration used in the empirical analysis.
(A) Mean county-level PM$_{2.5}$ concentrations.
(B) Spillover exposure induced by the Gaussian transport mapping.
(C) Gaussian transport network ($\sigma=25$ km) defining the corresponding exposure mapping.
}
\label{fig:transport_map}
\end{figure}

\begin{figure}[t]
\centering
\includegraphics[width=0.90\textwidth]{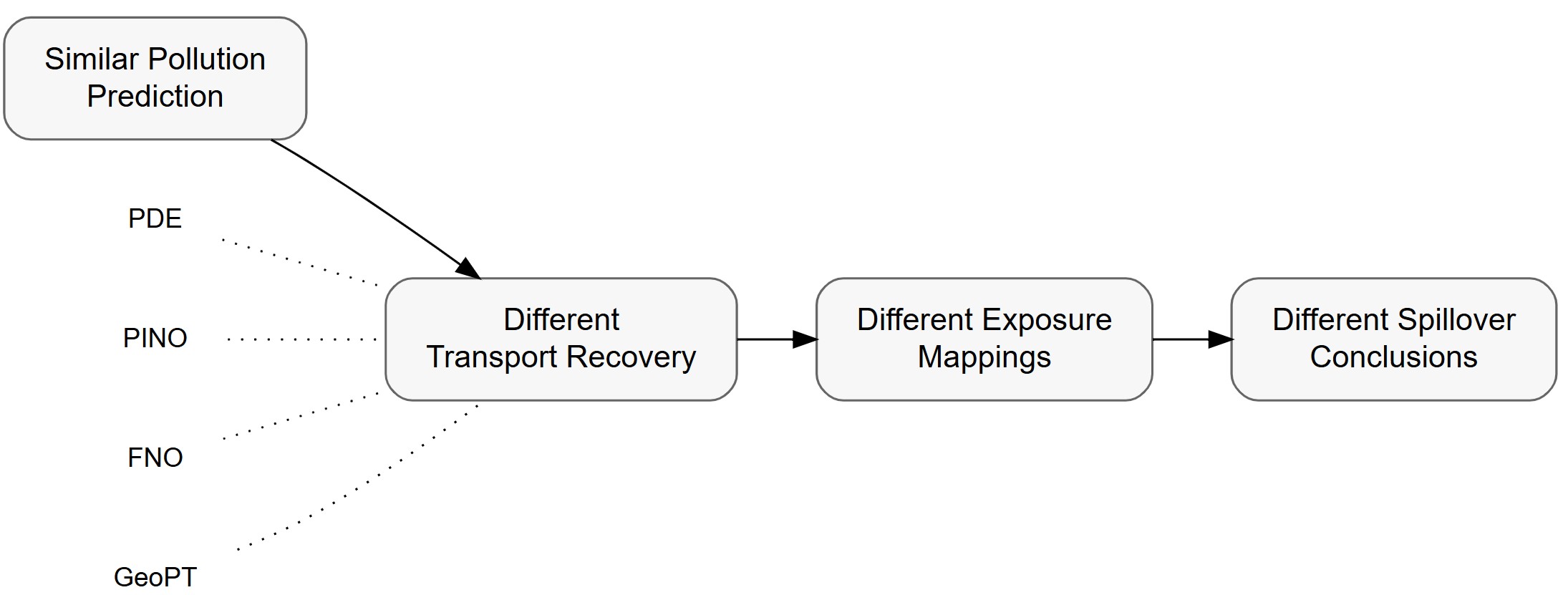}
\caption{
Summary of the proposed framework.
Transport models with similar prediction accuracy may recover different transport processes,
leading to different exposure mappings and ultimately different spillover conclusions.
}
\label{fig:summary}
\end{figure}

 Finally, we examine whether the same phenomenon appears in a real environmental application. We conclude with an illustration using California PM$_{2.5}$ data. Figure \ref{fig:transport_map} illustrates the empirical setting, including the observed PM$_{2.5}$ field, the induced spillover exposure, and the corresponding transport network used to construct the exposure mapping. The goal of this analysis was not to identify the true atmospheric transport process, which is unknown, but rather to examine how different transport assumptions affect inferred spillover patterns. After accounting for temporal persistence, transport effects remained positive across specifications and improved model fit. The transport-adjusted models consistently outperformed models based only on temporal information, suggesting that spatial transport contributed additional information beyond short-term persistence in PM$_{2.5}$ concentrations.  We then fit several competing transport models to the California data and examined the spillover patterns implied by each model under hypothetical pollution-control interventions. Although the competing transport models produced similar predictions of observed PM$_{2.5}$ concentrations, the locations identified as most affected by hypothetical pollution-control interventions differed across exposure mappings. The resulting spillover maps were not identical. Locations identified as highly affected under one transport model were not always identified as highly affected under another. Viewed together, the California analysis mirrors the simulation results. Different transport models can provide similar descriptions of the observed pollution field while implying different spillover patterns. Good prediction of pollution concentrations does not guarantee agreement in downstream spillover analyses. These empirical results illustrate that uncertainty in learned transport processes can propagate into downstream causal analyses even when competing models fit the observed pollution field similarly well.
 Although spillover estimates varied substantially across transport models, estimated intervention effects were considerably more stable. Across PDE, PINO, FNO, and GeoPT, estimated intervention effects ranged from 0.303 to 0.305 (Table~\ref{tab:policy_effects}). In this example, differences in the learned transport process had a much larger impact on spillover estimates than on the intervention effect itself. Transport models that produced noticeably different spillover estimates nevertheless yielded nearly identical estimates of the overall policy effect. Figure \ref{fig:summary} provides a graphical summary of this central finding.

\begin{table}[H]
\centering
\caption{Estimated intervention effects under competing transport models.}
\label{tab:policy_effects}
\begin{tabular}{lc}
\toprule
Model & Intervention Effect \\
\midrule
PDE   & 0.305 \\
PINO  & 0.303 \\
FNO   & 0.304 \\
GeoPT & 0.304 \\
\bottomrule
\end{tabular}
\end{table}

\section*{Discussion}
This paper studied how uncertainty in learned transport processes propagates through exposure mappings and ultimately affects causal spillover inference under interference. Although exposure mappings are often treated as known, in many applications they must be estimated from transport models and therefore inherit uncertainty from the learning process.
Our main finding is that similar predictive performance did not necessarily lead to similar spillover conclusions. Across PDE, PINO, FNO, and GeoPT, pollution prediction accuracy was nearly identical, yet estimated spillover effects varied substantially. The source of this discrepancy lay in the learned exposure mapping rather than in the prediction of the pollution field itself. Small differences in reconstructed exposure mappings translated into meaningful differences in spillover estimates, and exposure-mapping accuracy proved to be a stronger predictor of inferential performance than pollution prediction accuracy. These results suggest that predictive agreement alone is insufficient when learned transport models are used for causal inference.

A second finding is that the impact of exposure-mapping uncertainty depended on the scientific question being asked. Broad regional interventions produced relatively consistent conclusions across transport models, whereas localized point-source interventions showed substantially greater disagreement. Interestingly, this sensitivity was much more pronounced for spillover estimation than for policy evaluation: models that differed considerably in estimated spillover mechanisms often produced similar estimates of overall intervention effects. These results suggest that transport-model uncertainty may influence mechanistic understanding more than predictions of aggregate policy impact.

Although the California application cannot determine which transport model is correct, it illustrates that competing models with nearly identical predictive performance can imply different exposure mappings and consequently different spillover patterns. The empirical illustration therefore mirrors the simulation findings and demonstrates that these issues arise in realistic environmental settings rather than only in synthetic examples.

Several limitations should be noted. The simulations considered a relatively simple spatial setting and a limited set of transport models. In addition, the outcome model assumed a particular form of spillover exposure and did not consider more complex interference mechanisms. Future work could investigate richer intervention strategies, time-varying transport processes, uncertainty quantification for learned exposure mappings, and settings in which exposure mappings evolve.  

Beyond these specific limitations, our findings have broader implications for causal inference with learned scientific models. Transport models are commonly evaluated according to how well they reproduce observed pollution concentrations, yet models with nearly identical predictive performance can induce different exposure mappings and therefore support different causal conclusions. Our results suggest that evaluating transport models for causal applications should extend beyond predictive accuracy to include the stability and reliability of the induced exposure mappings.

While this paper focused on uncertainty arising from competing transport models, other sources of uncertainty—including measurement error, outcome-model misspecification, and errors unrelated to the transport process—remain important directions for future work.

Finally, our study suggests that exposure mappings should be viewed as estimated quantities rather than fixed design objects. Propagating uncertainty in learned exposure mappings through downstream analyses may therefore be essential for reliable causal inference under interference.
 \section*{Funding}

The author received no specific funding for this work.

\section*{Competing Interests}

The author declares no competing interests.

\section*{Data Availability}

The simulation data were generated by the author. The California PM$_{2.5}$ data used in the empirical illustration are publicly available. Processed data and code required to reproduce the analyses are available in the GitHub repository listed below.

\section*{Code Availability}

Code used to generate the simulations, analyses, and figures is publicly available at:

\url{https://github.com/congca/Causal-Inference-under-Interference-with-Learned-Exposure-Mappings}
 \bibliographystyle{plainnat}
\bibliography{references}

 \section*{Appendix}
 
Additional empirical results from the California PM2.5 analysis are summarized below. Past PM$_{2.5}$ levels already explained much of the short-term variation in pollution. However, after incorporating transport information such as wind-driven movement, the estimated relationships between regions changed noticeably. Across all models, the transport effects remained positive, suggesting that pollution spreads across nearby areas through spatial transport processes. Under the Gaussian transport mapping, the estimated spillover coefficient was:

\[
\hat{\beta}_{transport}
=
0.582
\quad
(p < 0.001)
\]

After adjusting for county-specific effects and overall time trends, the transport effect remained statistically significant. This indicates that pollution levels were still related across regions through spatial transport processes, even after accounting for temporal patterns over time. Residual Moran's I values were close to zero after including the transport adjustment, indicating that little spatial correlation remained unexplained in the residuals.

\[
Moran's\ I = -0.017
\]

These results imply that the transport model explained much of the remaining spatial correlation in the data. These findings motivated the simulation studies below, where we investigated how incorrect transport assumptions and overly broad spatial smoothing can affect spillover estimation and recovery of the underlying transport structure.

\begin{table}[H]
\centering
\caption{
Comparison of temporal persistence and transport-informed models.
}
\label{tab:main_models}

\begin{tabular}{lcccc}
\toprule
& Baseline & Spillover & County FE & Two-Way FE \\
\midrule

PM$_{2.5,t-1}$ 
& 0.667 
& 0.487 
& 0.217 
& 0.285 \\

& (0.036)
& (0.034)
& (0.027)
& (0.031)
\\

PM$_{2.5,t-2}$
& -0.043
& 0.021
& 
& 
\\

& (0.036)
& (0.032)
&
&
\\

Wind Speed
& -0.026
& -0.009
& -0.021
& -0.018
\\

& (0.006)
& (0.005)
& (0.005)
& (0.007)
\\

Temperature
& -0.006
& 0.009
& -0.079
& -0.368
\\

& (0.020)
& (0.017)
& (0.019)
& (0.050)
\\

Transport Operator
&
0.582
&
0.753
&
0.656
\\

&
(0.039)
&
(0.035)
&
(0.057)
\\

\midrule

Observations
& 799
& 799
& 799
& 799
\\

$R^2$
& 0.428
& 0.552
& 0.694
& 0.719
\\

RMSE
& 3.65
& 3.23
& 2.67
& 2.56
\\

AIC
& 4349.7
& 4156.8
& 3904.2
& 3891.9
\\

\bottomrule
\end{tabular}
\end{table}

\end{document}